\documentclass[aps,showpacs,superscriptaddress,twocolumn]{revtex4}

\usepackage{amsfonts}
\usepackage{amsmath}
\usepackage{amssymb}
\usepackage{graphicx}

\input{tcilatex}

\begin{document}

\title{Optical properties of UO$_{2}$ and PuO$_{2}$}
\author{Hongliang Shi}
\affiliation{LCP, Institute of Applied Physics and Computational Mathematics, P.O. Box
8009, Beijing 100088, People's Republic of China}
\affiliation{SKLSM, Institute of Semiconductors, Chinese Academy of Sciences, P. O. Box
912, Beijing 100083, People's Republic of China}
\author{Mingfu Chu}
\affiliation{State Key Laboratory for Surface Physics and Chemistry, Mianyang 621907,
People's Republic of China}
\author{Ping Zhang}
\thanks{Author to whom correspondence should be addressed. Electronic
address: zhang\_ping@iapcm.ac.cn}
\affiliation{LCP, Institute of Applied Physics and Computational Mathematics, P.O. Box
8009, Beijing 100088, People's Republic of China}
\affiliation{Center for Applied Physics and Technology, Peking University, Beijing
100871, People's Republic of China}

\begin{abstract}
We perform first-principles calculations of electronic structure and
optical properties for UO$_{2}$ and PuO$_{2}$ based on the density
functional theory using the generalized gradient approximation
(GGA)+\emph{U} scheme. The main features in orbital-resolved partial
density of states for occupied \emph{f} and \emph{p} orbitals,
unoccupied \emph{d} orbitals, and related gaps are well reproduced
compared to experimental observations. Based on the satisfactory
ground-state electronic structure calculations, the dynamical
dielectric function and related optical spectra, i.e., the
reflectivity, adsorption coefficient, energy-loss, and refractive
index spectrum, are obtained. These results are consistent well with
the attainable experiments.
\end{abstract}

\maketitle

\section{INTRODUCTION}

Actinide dioxides (AnO$_{2}$) have been attracted lots of attention due to
their rich physical phenomena characterized by the complex nature of 5\emph{f%
} electrons. Many experimental and theoretical works have been
devoted to investigating the thermodynamical, electronic structural,
and defect properties of AnO$_{2}$ systems. Taking UO$_{2}$ and
PuO$_{2}$ for example, their insulating ground states have been
established experimentally \cite{r1,r2} and successfully predicted
theoretically \cite{r3,r4,r5,r6}. When referring to insulators or
semiconductors, one basic physical quantity of interest is their
band gaps. If the band gap of UO$_{2}$ or PuO$_{2}$ can be
comparable to semiconductors, one idea may occur to us that whether
they can be applied extensively in the electronic and optoelectronic
devices like semiconductors (Si, GaAs, and ZnO) or not. Recently,
Meek \emph{et al.} discussed the electronic properties of uranium
dioxide and revealed the potential performance advantages of uranium
dioxide as compared to conventional semiconductor materials
\cite{r7}. Especially, the higher dielectric constant of UO$_{2}$
makes it more suitable for making integrated circuits \cite{r7}.
This may stimulate many studies of the optical properties for
actinide dioxides in future.

Optical adsorption and reflectance spectra of semiconductors have been
studied for several decades both experimentally and theoretically, whereas,
similar works performed on actinide dioxides is still very scarce although
they are necessary not only from the viewpoint of basic science but also
from their technological importance in industries. Experimentally, Schoenes
studied the incidence reflectivity of UO$_{2}$ single crystals in the photon
energy range of 0.03-13 eV, from which the complex dielectric function $%
\varepsilon(\omega)=\varepsilon_{1}(\omega)+
i\varepsilon_{2}(\omega)$ has been derived \cite{r8}. For PuO$_{2}$,
to our knowledge, no experimental optical data are available in
literature. As for the theoretical investigations of optical
spectrum of actinide dioxides, it is a great challenge to standard
density functional theory that an accurate description of electronic
structure for actinide oxides is hard to be achieved, which is
indispensable to getting the correct optical spectrum. Conventional
density functional schemes that apply the local density
approximation (LDA) or the GGA underestimate the strong on-site
Coulomb repulsion of the 5$f$ electrons and consequently fail to
capture the correlation-driven localization. Therefore, the 5$f$
electrons in actinide oxides require special attention. One
promising way to improve contemporary LDA and GGA approaches is to
modify the intra-atomic Coulomb interaction through the so-called
LDA+$U$ or GGA+$U$ approach, in which the underestimation of the
intraband Coulomb interaction is corrected by the Hubbard $U$
parameter \cite{Ani1993,Sol1994}. Recently, the electronic
structures of UO$_{2}$ and PuO$_{2}$ are correctly reproduced using
LDA+\emph{U} or GGA+\emph{U} calculations \cite{r3,r4,r5,r6}.
Therefore, based on the good performance of LDA/GGA+\emph{U}
approaches in describing the electronic structure of the systems
containing 5\emph{f} electrons, it is encouraging to investigate the
optical spectra of them.

In this work, we used the GGA+\emph{U} scheme to study the static
and frequency-dependent dynamical dielectric response functions for
UO$_{2}$ and PuO$_{2}$. Our present calculated band gap $E_{g}$ and
high-frequency dielectric constant $\varepsilon_{\infty}$ for
UO$_{2}$ are 2.3 eV and 5.53, which are in good agrement with the
experimental values of about 2.1 eV and 5.1 observed in the optical
spectra \cite{r8}, respectively. Furthermore, our calculated
dielectric function $\varepsilon(\omega)$ exhibits the overall
agreement with experimental result and the main peaks are well
reproduced. The dielectric function and the consequent optical
spectra for
PuO$_{2}$ are also calculated in the paper. In particular, the value of $%
\varepsilon_{\infty}$ for PuO$_{2}$ is predicted to be 6.21, a
little larger than that for UO$_{2}$. Considering the satisfactory
calculations for UO$_{2} $, we expect our predicted optical behavior
for PuO$_{2}$ can provide a useful reference for future experimental
measurement.

\section{DETAILS OF CALCULATION}

Our electronic structural and optical calculations are performed
using the projector-augmented wave (PAW) method of Bl\"{o}chl
\cite{r111}, as implemented in the ab initio total-energy and
molecular-dynamics program VASP (Vienna \textit{ab initio}
simulation program) \cite{r13}. PAW is an all-electron method that
combines the accuracy of augmented-plane-wave methods with the
efficiency of the pseudopotential approach. The PAW method is
implemented in VASP with the frozen-core approximation. The
exchange-correlation functional is used GGA of
Perdew-Burke-Ernzerhof (PBE) formalism \cite{r14}. The 5\emph{f},
6\emph{s}, 6\emph{p}, 6\emph{d}
and 7\emph{s} electrons of U and Pu as well as the oxygen 2\emph{s} and 2%
\emph{p} electrons are explicitly treated as valence electrons. The electron
wave function is expanded in plane waves up to a cutoff energy of 500 eV.
For the Brillouin zone integration, the $\Gamma$ centered 6$\times$6$\times$%
6 grid is adopted. 144 bands are used to get the dynamical
dielectric function $\varepsilon(\omega)$ and a good convergence can
be achieved. In order to perform the antiferromagnetic (AFM) phase
calculations, we used the unit cell containing 12 atoms. The strong
on-site Coulomb repulsion among the localized 5\emph{f} electrons is
described by using the formalism formulated by Dudarev \emph{et al.
}\cite{r15}. In this scheme, only the difference between the
spherically averaged screened Coulomb energy \emph{U} and the
exchange energy \emph{J} is important for the total LDA (GGA) energy
functional. Thus, in the following we label them as one single
effective parameter \emph{U} for brevity. In our calculation, we use
\emph{J}=0.51 and 0.75 eV for the exchange energies of U and Pu,
respectively, and the effective Hubbard \emph{U} are 4.0 and 3 eV,
which are close to the values used in other previous work
\cite{r4,r5}.

As for the optical spectra calculations, we adopt two different
methods to determine the macroscopic static dielectric constants
using different approximations \cite{r17}. One method is using a
summation over conduction band states and the other is using the
linear response theory (density functional theory). For the latter,
only the static ion-clamped dielectric matrix can be obtained and a
summation over empty conduction band states is not required, whereas
the former can calculate the frequency-dependent dynamic dielectric
function after the electronic ground state has been obtained. The
frequency-dependent imaginary part of the dielectric function is
determined by a summation over empty states using the following
equation \cite{r17}:
\begin{eqnarray}
\varepsilon _{\alpha \beta }^{(2)}(\omega ) &=&\frac{4\pi ^{2}e^{2}}{\Omega }%
\lim_{q\rightarrow 0}\frac{1}{q^{2}}\sum_{c,v,\mathbf{k}}2w_{\mathbf{k}%
}\delta (\varepsilon _{c\mathbf{k}}-\varepsilon _{v\mathbf{k}}-\omega )
\notag \\
&&\times \left\langle u_{c\mathbf{k}+\mathbf{e}_{\alpha }\mathbf{q}%
}|u_{vk}\right\rangle \left\langle u_{c\mathbf{k}+\mathbf{e}_{\beta }\mathbf{%
q}}|u_{v\mathbf{k}}\right\rangle ^{\ast },
\end{eqnarray}%
where the indices $c$ and $v$ refer to conduction and valence band
states respectively, and $u_{c\mathbf{k}}$ is the cell periodic part
of the wavefunctions at the \emph{k}-point \textbf{k}. The real part
of the dielectric tensor is obtained by the usual Kramers-Kronig
transformation

\begin{equation}
\varepsilon _{\alpha \beta }^{(1)}(\omega )=1+\frac{2}{\pi }%
P\int_{0}^{\infty }\frac{\varepsilon _{\alpha \beta }^{(2)}(\omega
^{^{\prime }})\omega ^{^{\prime }}}{\omega ^{^{\prime }2}-\omega ^{2}+i\eta }%
d\omega ^{^{\prime }},
\end{equation}
where \emph{P} denotes the principle value.

The main optical spectra, such as the reflectivity \emph{R}($\omega
$),
adsorption coefficient \emph{I}($\omega $), energy-loss spectrum \emph{L}($%
\omega $), and refractive index \emph{n}($\omega $), all can be obtained
from the dynamical dielectric response functions $\varepsilon (\omega )$.
The explicit expressions are given by
\begin{equation}
R(\omega )=\left\vert \frac{\sqrt{\varepsilon (\omega )}-1}{\sqrt{%
\varepsilon (\omega )}+1}\right\vert ^{2},
\end{equation}%
\begin{equation}
I(\omega )=(\sqrt{2})\omega \left[ \sqrt{\varepsilon _{1}(\omega
)^{2}+\varepsilon _{2}(\omega )^{2}}-\varepsilon _{1}(\omega )\right] ^{1/2},
\end{equation}

\begin{equation}
L(\omega )=\varepsilon _{2}(\omega )/\left[ \varepsilon _{1}(\omega
)^{2}+\varepsilon _{2}(\omega )^{2}\right],
\end{equation}
and
\begin{equation}
n(\omega )=(1/\sqrt{2})\left[ \sqrt{\varepsilon _{1}(\omega
)^{2}+\varepsilon _{2}(\omega )^{2}}+\varepsilon _{1}(\omega )\right] ^{1/2},
\end{equation}
respectively.

\section{results and discussions}

\subsection{electronic structure and optical properties of UO$_{2}$}

\begin{table}[tbp]
\caption{Ion clamped static macroscopic dielectric constants $\protect%
\varepsilon_{\infty}$ of UO$_{2}$ and PuO$_{2}$ calculated using the PAW
method and various approximations with various \emph{k}-points sampling: $%
\Gamma$ indicates a grid centered at $\Gamma$ point, whereas
Monkhorst-Pack (MP) grids do not contain the $\Gamma$ point. N$_{k}$
stands the number of
irreducible \emph{k}-points of the Brillouin zone (IBZ) at specific \emph{k}%
-points sampling. $\protect\varepsilon_{\mathrm{mic}}$ indicates values
neglecting local field effects, $\protect\varepsilon_{\mathrm{RPA}}$
includes local fields effects in the Hartree approximation, and $\protect%
\varepsilon_{\mathrm{DFT}}$ includes local fields effects on the DFT level. $%
\protect\varepsilon^{\mathrm{cond}}$ are values obtained by summation over
conduction band states, whereas $\protect\varepsilon^{\mathrm{LR}}$ are
values obtained using linear response theory (density functional
perturbation theory).}
\label{tab:table1}%
\begin{ruledtabular}
\begin{tabular}{cccccccccc}
AnO$_{2}$ &\emph{k}-mesh &N$_{k}$(IBZ) & $\varepsilon^{\rm{LR}}_{\rm{mic}}$ & $\varepsilon^{\rm{LR}}_{\rm{RPA}}$ & $\varepsilon^{\rm{LR}}_{\rm{DFT}}$ & $\varepsilon^{\rm{cond}}_{\rm{mic}}$\\
\hline
UO$_{2}$ & &&& & \\
         &(12$\times$12$\times$12)$\Gamma$ &196 &5.71&5.28&5.53&5.59 \\
         &(8$\times$8$\times$8)$\Gamma$ &75&5.71&5.28 &5.53& 5.59\\
         &(6$\times$6$\times$6)$\Gamma$ &40 &5.71&5.28&5.53& 5.59\\
         &(6$\times$6$\times$6)MP       &18&5.71&5.28&5.53& 5.59\\

PuO$_{2}$ &&&&& \\
         &(12$\times$12$\times$12)$\Gamma$&196&6.38&5.94&6.21& 6.23\\
         &(8$\times$8$\times$8)$\Gamma$   &75&6.37&5.94&6.20&6.23 \\
         &(6$\times$6$\times$6)$\Gamma$   &40&6.37&5.94&6.21&6.23 \\
         &(6$\times$6$\times$6)MP         &18&6.37&5.94&6.20&6.23 \\
\end{tabular}
\end{ruledtabular}
\end{table}

\begin{figure}[ptb]
\includegraphics*[height=11.6cm,keepaspectratio]{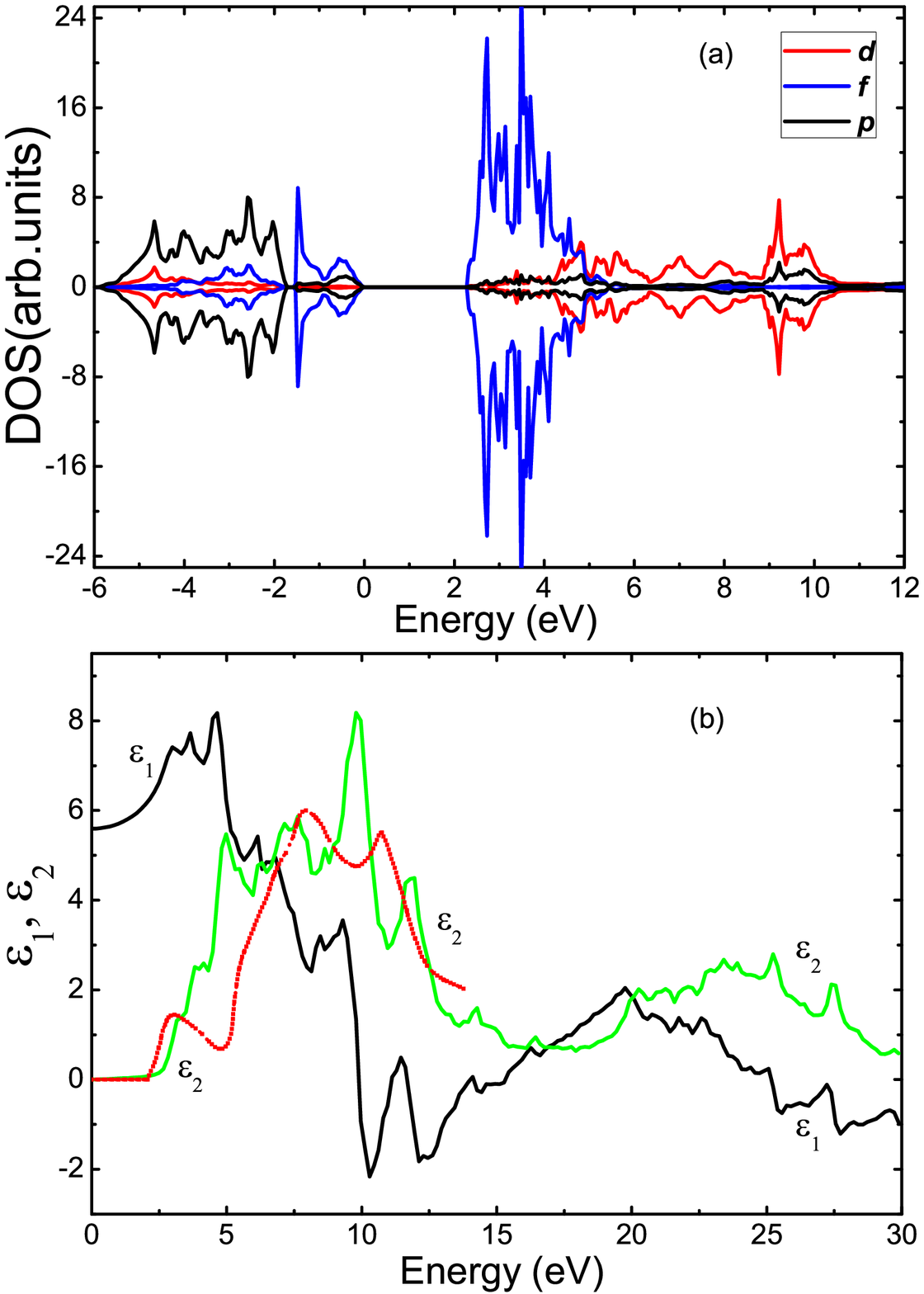}
\caption{(a) The projected orbital-resolved partial DOS for U 6%
\emph{d}, U 5\emph{f}, and O 2\emph{p} orbitals in antiferromagnetic
UO$_{2}$. The Fermi level is set to
zero. (b) The dynamical dielectric function $\protect\varepsilon(%
\protect\omega)=\protect\varepsilon_{1}(\protect\omega)+ \emph{i}\protect\varepsilon%
_{2}(\protect\omega)$ as a function of the photon energy $\omega$
for UO$_{2}$. The
black and green lines represent our calculated real and imaginary parts of dielectric function $%
\protect\varepsilon(\protect\omega)$, respectively, while the red
dotted-line is experimental $\protect\varepsilon_{2}(\protect\omega)$.}
\label{fig1:epsart}
\end{figure}

\begin{figure}[ptb]
\includegraphics*[height=7.0cm,keepaspectratio]{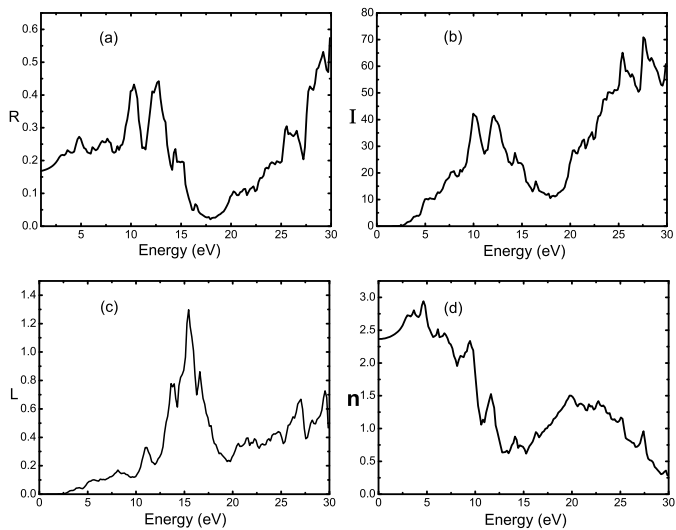}
\caption{Calculated optical spectra for UO$_{2}$, (a) the reflectivity \emph{%
R}($\protect\omega $), (b) adsorption coefficient \emph{I}($\protect\omega $%
), (c) energy-loss \emph{L}($\protect\omega $), and (d) refractive index
\emph{n}($\protect\omega $).}
\label{fig2:epsart}
\end{figure}

Since the optical spectra are directly calculated from interband
transitions, an accurate description of the electronic structure is
indispensable. The calculated orbital-resolved partial density of
states (PDOS) for U 5\emph{f}, U 6\emph{d} and O 2\emph{p} are shown
in Fig. 1(a). The Fermi level is set to be zero. It is clearly shown
that the valence bands are mainly contributed by U 5\emph{f} and O
2\emph{p}
orbitals. The peak near the Fermi level is mainly U 5\emph{f} with a little O 2%
\emph{p} contribution, which bas been confirmed by the resonant photoemission%
 \cite{r18}. The U 5\emph{f} valence band covers from 0 to $-$1.6 eV, which is
also consistent with the experimental observation that the occupied
5\emph{f} states in UO$_{2}$ are located around 1.5 eV below the
Fermi level with a band width of about 2.0 eV \cite{r18}. The O
2\emph{p} valence band width is 4.0 eV from about $-$1.8 to $-$5.8
eV, in qualitative agreement with the photoemission value of 5.0 eV
from $-$3.0 to from $-$8.0 eV \cite{r18}.

As for the unoccupied U 5\emph{f} and 6\emph{d} orbitals, their
accurate descriptions are also indispensable to the interband
transitions, since electrons are excited from the occupied valence
bands to the unoccupied bands during optical excitations. The
5\emph{f} and 6\emph{d} bands begin at about 2.3 and 4 eV,
respectively, which are consistent well with the results of 2.6 and
5 eV obtained by hybrid DFT method \cite{r19}. Note that our
calculated \emph{p} $\rightarrow$ \emph{d} gap is 5.8 eV, which
accords well with the
Bremsstrahlung Isochromat Spectroscopy (BIS) value of 5.0$\pm$0.4 eV \cite%
{r20}. Overall, our calculated DOS agrees well the experimental
spectra and other theoretical results. This supplies the safeguard
for our following optical spectrum calculations.

Due to the cubic symmetry of UO$_{2}$, the dielectric tensor only has one
independent component and $\varepsilon_{xx}$=$\varepsilon_{yy}$=$%
\varepsilon_{zz}$. Our calculated macroscopic dielectric constants $%
\varepsilon_{\infty}$ using different methods and approximations are
collected in Table I. We find that well converged results can be
obtained by using the $\Gamma$-centered 6$\times$6$\times$6 grid.
Note that the value of $\varepsilon^{\mathrm{LR}}_{\mathrm{DFT}}$
should be compared to experiment. For UO$_{2}$, the calculated
$\varepsilon_{\infty}$ is 5.53, which agrees well with the
experimental value of 5.1 \cite{r8}.

As for the dynamical dielectric function, our calculated imaginary part $%
\varepsilon_{2}(\omega)$ and real part $\varepsilon_{1}(\omega)$ of
the complex dielectric function $\varepsilon(\omega)$ together with
the corresponding experimental $\varepsilon_{2}(\omega)$ are showed
in Fig. 1(b). The green and black lines represent our calculated
imaginary and real parts of the complex dielectric function
$\varepsilon(\omega)$, respectively, while the red dotted-line gives
the experimental measurement \cite%
{r8} of $\varepsilon_{2}(\omega)$. Our theoretical photon energy
covers from 0 to 30 eV, while the experimental \cite{r8} value
covers from 0 to 13 eV. According to our calculated DOS showed in
Fig. 1(a), we suggest that in $\varepsilon_{2}(\omega)$ the peaks
(at 2.8 eV) below 3 eV should be assigned to the intra 5\emph{f}
transitions. Notice that the unoccupied 6\emph{d} bands begin about
at 4 eV, therefore, the 5\emph{f} $\rightarrow$ 6\emph{d} transition
energies should be larger than 4 eV. Kudin \emph{et al.} also
suggested that the stronger adsorption observed experimentally at
$\sim$5-6 eV could be assigned to the optically allowed 5\emph{f}
$\rightarrow$ 6\emph{d}
transitions \cite{r19}. According to our calculated $\varepsilon_{2}(\omega)$%
, four main peaks lie at about 5.0, 7.1, 9.8, and 11.8 eV,
respectively. The shape of the calculated curve exhibits the same
main features demonstrated by the experimental results \cite{r8}.
Combined with the orbital-resolved PDOS shown in Fig. 1(a), we
attribute the first two peaks in $\varepsilon_{2}(\omega)$ to be
5\emph{f} $\rightarrow$ 6\emph{d} transitions, while the last two to
be 2\emph{p} $\rightarrow$ 6\emph{d} transitions. This is consistent
well with the experimental assignment \cite{r21} by Naegele \emph{et
al.}, who attributed the peak around 3 eV in
$\varepsilon_{2}(\omega)$ to intra 5\emph{f}$^{2}$ transitions,
while the peak structures above 5 and 10 eV were ascribed to the
\emph{f} $\rightarrow$ \emph{d} and \emph{p} $\rightarrow$ \emph{d}
transitions, respectively. Another assignment was suggested by
Schoenes according to their dielectric function deduced from the
reflectivity measurement; they argued that the peaks near 3 and 6 eV
correspond to \emph{f} $\rightarrow$ \emph{d} transitions, and that
the peaks near 8 and 11 eV are due to \emph{p} $\rightarrow $
\emph{d} transitions \cite{r8,r23}. Herein, the assignment of
\emph{f} $\rightarrow$ \emph{d} transition at 3 eV in Ref.
\cite{r8,r23} is not supported by our calculation. The cause is that
in assigning the peak in $\varepsilon_{2}(\omega)$ at 3 eV, the
energy distance between U occupied 5$f^{2}$ and O 2$p$ valence bands
was overestimated in Ref. \cite{r8,r23} to be as large as 4 eV,
which is much larger than that directly determined by the
photoemission measurements \cite{r18,r24,r25}. On the contrary,
according to our band-structure calculation, the occupied 5\emph{f}
orbitals are locate at about 1.5 eV below the Fermi level and the O
2$p$ bands widely covers from about -1.8 to -5.8 eV, which instead
accords well with the experimental photoemission data
\cite{r18,r24,r25} in UO$_{2}$. Thus, as mentioned above, we suggest
the structure in $\varepsilon_{2}(\omega)$ below 3 eV is caused by
the intra 5\emph{f} transitions.

Using expressions (3)-(7), the reflectivity \emph{R}($\omega $),
adsorption coefficient \emph{I}($\omega $), energy-loss
\emph{L}($\omega $) and refractive index \emph{n}($\omega $) spectra
are showed in Fig. 2. For reflectivity \emph{R}($\omega $) spectrum,
there are four peaks locating at 4.8, 7.6, 10.3, and 12.8 eV. The
adsorption coefficient \emph{I}($\omega $) spectrum has the same
trends. The origin of these peaks can also be explained as the peaks
of the imaginary part $\varepsilon_{2}(\omega)$. Note that three
similar peaks at 5.5, 8, 11.7 eV are also observed by the
reflectance spectrum up to 13 eV at room temperature for UO$_{2}$
\cite{r8}. The energy-loss \emph{L}($\omega $) spectrum can
demonstrate not only one-particle excitations but also collective
excitations. The maxima at around 15.4 eV as showed in Fig. 2(c)
indicates the plasmon resonance, which is qualitatively consistent
with the experimental value of 14 eV \cite{r23}. As showed in Fig.
1(b), at about 11.7 eV the real part $\varepsilon_{1}$ becomes zero,
arriving at the minima around 12.1 eV and then approaches zero at
about 14 eV. As Schoenes pointed out, the energy at which
$\varepsilon_{1}(\omega)$ crosses the zero line with a positive slop
gives the plasmon excitation energy \cite{r23}.

\subsection{electronic structure and optical properties of PuO$_{2}$}

\begin{figure}[ptb]
\includegraphics*[height=11.6cm,keepaspectratio]{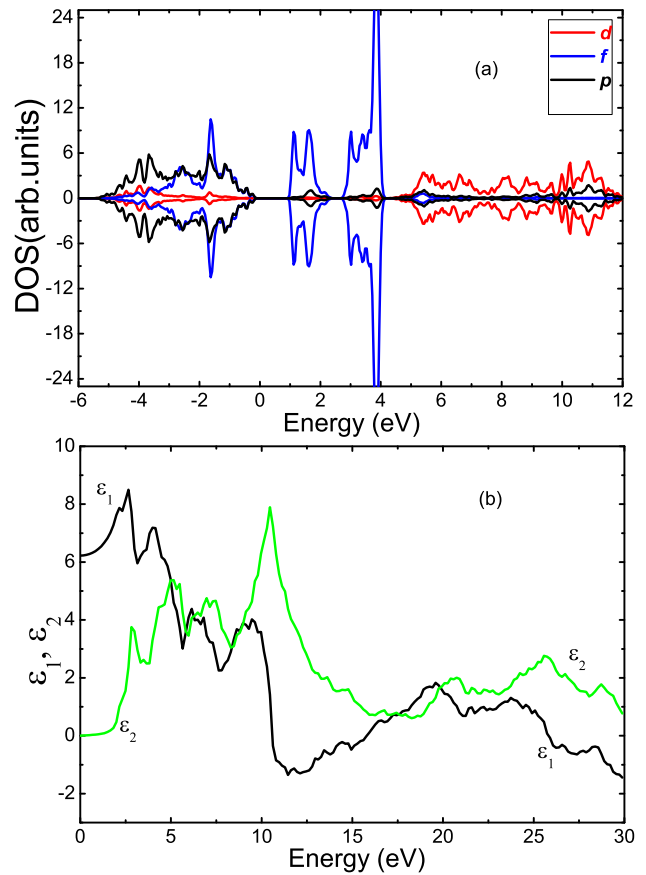}
\caption{(a) The projected orbital-resolved partial DOS for Pu 6%
\emph{d}, Pu 5\emph{f}, and O 2\emph{p} orbitals in
antiferromagnetic PuO$_{2}$. The Fermi level is set to
zero. (b) The dynamical dielectric function $\protect\varepsilon(%
\protect\omega)=\protect\varepsilon_{1}(\protect\omega)+ \emph{i}\protect\varepsilon%
_{2}(\protect\omega)$ as a function of the photon energy $\omega$
for PuO$_{2}$. The
black and green lines represent our calculated real and imaginary parts of dielectric function $%
\protect\varepsilon(\protect\omega)$, respectively.}
\label{fig3:epsart}
\end{figure}

Due to Pu unique position of its 5\emph{f} electrons between
localized and delocalized states among actinide series, Pu metal and
plutonium-based oxides have more complex properties than other
actinides. For example, metallic Pu has six different phase under
different temperatures and pressures \cite{r26}. PuO$_{2}$ as an
important actinide dioxide has extensive applications in nuclear
reactor fuel and long-term storage of surplus plutonium. Therefore,
the study of optical properties for PuO$_{2}$ is also necessary and
interesting. However, no experimental results of optical properties
for PuO$_{2} $ in literature are available. Recently, Butterfield
\emph{et al.} studied the photoemission behavior of surface oxides
of $\delta$-plutonium and they observed that two peaks
characterized by Pu 5\emph{f} and O 2\emph{p} orbitals are dominant in PuO$%
_{2}$ and Pu$_{2}$O$_{3}$ \cite{r27}. For PuO$_{2}$, the two peaks
observed are located at approximately 2.5 and 4.6 eV \cite{r27}, and
our calculated DOS showed in Fig. 3(a) also present two similar
peaks, i.e., a strong peak at about 1.6 eV and a weaker one at 3.7
eV. Overall, these features are well reflected in our PDOS showed in
Fig. 3(a) compared to experimental observations. As for the
unoccupied 6\emph{d} states, no experimental data can be obtained.
Our calculated unoccupied 6\emph{d} states begin at about 5 eV.
Considering the O 2\emph{p} peak at $-$3.7 eV, we suggest the
\emph{p} $\rightarrow$ \emph{d} transitions occur at larger than 9
eV.

Our calculated macroscopic dielectric constant
$\varepsilon_{\infty}$ for PuO$_{2}$ are also collected in Table I.
The present $\varepsilon_{\infty}$ is 6.21, whereas, no experimental
value is unavailable at present. Our calculated imaginary part $%
\varepsilon_{2}(\omega)$ and real part $\varepsilon_{1}(\omega)$ of
the complex dielectric function $\varepsilon(\omega)$ are showed in
Fig. 3(b). For $\varepsilon_{2}(\omega)$, four main peaks locate at
2.8, 5.1, 7.5, and 10.5 eV. According to our PDOS calculation showed
in Fig. 3(a), we attribute the peak below 6 eV to be intra 5\emph{f}
transitions, and the last two to be \emph{f} $\rightarrow$ \emph{d}
and \emph{p} $\rightarrow$ \emph{d} transitions, respectively. The
two similar peaks at 7 and 10 eV are also
obtained by Jomard \emph{et al.} using \emph{ab initio} calculations \cite%
{r28}.

Other related optical spectra for PuO$_{2}$ are showed in Fig. 4.
For reflectivity \emph{R}($\omega $) spectrum, there are four peaks
at 2.8, 5.0, 7.5, and 10.6 eV. Similarly, four peaks at 3.0, 5.5,
7.6, 10.6 eV are also observed in the adsorption coefficient
\emph{I}($\omega $) spectrum. The origin of these peaks can also be
explained according to the structure displayed in the imaginary part
$\varepsilon_{2}(\omega)$ of the dielectric function. It is evident
that the plasmon excitation occurs at 16.0 eV, which is similar to
the case of UO$_{2}$ at 15.4 eV as mentioned above.

\begin{figure}[ptb]
\includegraphics*[height=6.8cm,keepaspectratio]{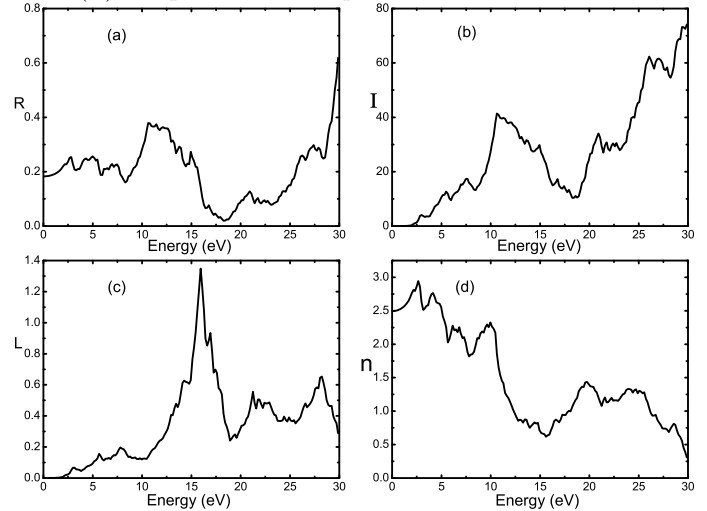}
\caption{Calculated optical spectra for PuO$_{2}$, (a) the reflectivity
\emph{R}($\protect\omega $), (b) adsorption coefficient \emph{I}($\protect%
\omega $), (c) energy-loss \emph{L}($\protect\omega $), and (d) refractive
index \emph{n}($\protect\omega $).}
\label{fig4:epsart}
\end{figure}

\section{SUMMARY}

In summary, we have performed a detailed investigation of the
electronic structure and optical spectra of actinide dioxides
UO$_{2}$ and PuO$_{2}$ using first-principle methods. For UO$_{2}$,
our calculated projected orbital-resolved PDOS for U 5\emph{f} and O
2\emph{p} orbitals in the valence region agree well with the
experimental photoemission observation. As for the unoccupied
states, our calculated \emph{p-d} gap is 5.8 eV, similar to the
experimental BIS value of 5.0$\pm$0.4 eV. The calculated
insulating band gap $E_{g}$ and macroscopic static dielectric constants $%
\varepsilon_{\infty}$ for UO$_{2}$ are 2.3 eV and 5.53,
respectively, which are also in good agrement with the experimental
values of about 2.1 eV and 5.1. The main features in spectra for
UO$_{2}$ are also well reproduced by our calculated dynamical
dielectric function $\varepsilon(\omega)$ compared to the
experimental observation. For PuO$_{2}$, the two main peaks
characterized by Pu 5\emph{f} and O 2\emph{p} orbitals in valence
bands are evidenced in our calculated PDOS, which accords well with
the photoemission results. The calculated macroscopic static
dielectric constants $\varepsilon_{\infty}$ is 6.21. The related
optical spectra for PuO$_{2}$ are also obtained by calculating the
dynamical dielectric function. The \emph{f} $\rightarrow$ \emph{d}
and \emph{p} $\rightarrow$ \emph{d} transitions are found to occurr
at 7.5 and 10.5 eV, respectively. Considering the satisfactory
optical description for UO$_{2}$ compared to experiments, we expect
that these results for PuO$_{2}$ are also reasonable and therefore
can provide a useful reference for future experimental measurement.

\begin{acknowledgments}
This work was supported by the Foundations for Development of Science and
Technology of China Academy of Engineering Physics under Grant No.
2009B0301037.
\end{acknowledgments}


\begin{references}

\bibitem{r1}
J. Faber, Jr., G. H. Lander, and B. R. Cooper, Phys. Rev. Lett.
\textbf{35} 1770 (1975).
\bibitem{r2}
C. E. McNeilly, J. Nucl. Mater. \textbf{11}, 53 (1964).
\bibitem{r3}
H. Geng, Y. Chen, Y. Kaneta, and M. Kinoshita, Phys. Rev. B
\textbf{77}, 180101 (2008).
\bibitem{r4}
B. Sun, P. Zhang, and X.-G. Zhao, J. Chem. Phys. \textbf{128},
084705 (2008).
\bibitem{r5}
D. A. Andersson, J. Lezama, B. P. Uberuaga, C. Deo, and S. D.
Conradson, Phys. Rev. B \textbf{79}, 024110 (2009).
\bibitem{r6}
L. Petit, A. Svane, Z. Szotek, W. M. Temmerman, and G. M. Stocks,
arXiv:0908.1806v1
\bibitem{r7}
T. Meek, M. Hu, and M. Haire,
http://www.nuenergy.org/pdf/UO2semicond.pdf
\bibitem{r8}
J. Schoenes, J. Appl. Phys. \textbf{49}, 1463 (1978).


\bibitem {Ani1993}V.I. Anisimov, I.V. Solovyev, M.A. Korotin, M.T. Czy\.{z}yk,
and G.A. Sawatzky, Phys. Rev. B \textbf{48}, 16929 (1993).

\bibitem {Sol1994}I.V. Solovyev, P.H. Dederichs, and V.I. Anisimov, Phys. Rev.
B \textbf{50}, 16861 (1994).


\bibitem{r111}
P.E.Bl$\ddot{\rm{o}}$chl, Phys. Rev. B 50, 17953 (1994).
\bibitem{r13}
G. Kresse and J. Hafner, Phys. Rev. B \textbf{48}, 13115 (1993).

\bibitem{r14}
J. P. Perdew, K. Burke, and M. Ernzerhof, Phys. Rev. Lett. 77, 3865
(1996).

\bibitem{r15}
S. L. Dudarev, G. A. Botton, S. Y. Savrasov, C. J. Humphreys, and A.
P. Sutton, Phys. Rev. B \textbf{57}, 1505 (1998).
\bibitem{r17}
M. Gajdo$\check{\rm{s}}$, K. Hummer, G. Kresse, J.
Furthm$\ddot{\rm{u}}$ller, and F. Bechstedt, Phys. Rev. B
\textbf{73}, 045112 (2006).

\bibitem{r18}
L. E. Cox, W. P. Ellis, R. D. Cowan, J.W. Allen, S.-J. Oh, I.
Lindau, B. B. Pate, and A. J. Arko, Phys. Rev. B \textbf{35}, 5761
(1987).

\bibitem{r19}
K. N. Kudin, G. E. Scuseria, and R. L. Martin, Phys. Rev. Lett.
\textbf{89}, 266402 (2002).

\bibitem{r20}
Y. Baer and J. Schoenes, Solid State Commun. \textbf{33}, 885
(1980).
\bibitem{r21}
J. Naegele, L. Manes and U. Birkholz, Proc. 5th Int. Conf. Plutonium
and other Actinides, ed. H.Blank and R. Lidner (North-Holland,
Amsterdam, 1976), p. 393.

\bibitem{r23}
J. Schoenes, Phys. Rep. \textbf{63}, 301 (1980).
\bibitem{r24}
J. Naegele, J. de Phys. C4, 169 (1979).
\bibitem{r25}
P.R. Norton, R.L. Tapping, D.K. Creber and W.J.L. Buyers, Phys. Rev.
B \textbf{21}, 2572 (1980).
\bibitem {r26}
K. T. Moore and G. van der Laan, Rev. Mod. Phys. \textbf{81}, 235
(2009).
\bibitem {r27}
M. Butterfield, T. Durakiewicz, E. Guziewicz, J. Joyce, A. Arko,
K. Graham, D. Moore, and L. Morales, Surf. Sci. \textbf{571}, 74
(2004).
\bibitem {r28}
G. Jomard, B. Amadon, F. Bottin, and M. Torrent, Phys. Rev. B
\textbf{78}, 075125 (2008).


\end{references}
\end{document}